\documentclass[12pt]{article} 
\usepackage[dvips]{graphics}
\title{Rotation Representations and e, $\pi$, p Masses}  
\author{{\it Richard Shurtleff~}\thanks{affiliation and mailing 
address: Department of Mathematics and Applied Sciences, 
Wentworth Institute of Technology, 550 Huntington Avenue, 
Boston, MA, USA, ZIP 02115, telephone number: (617) 989-4338, fax 
number: (617) 989-4591 , e-mail address: shurtleffr@wit.edu}} 
\date{July 2, 1999}
\begin{document} 
          
\maketitle               
			\begin{abstract} 
Mass is proportional to phase gain per unit time; for e, $\pi$, and p the quantum frequencies are 0.124, 32.6, and 227 Zhz, respectively. By explaining how these particles acquire phase at different rates, we explain why these particles have different masses. Any free particle spin 1/2 wave function is a sum of plane waves with spin parallel to velocity. Each plane wave, a pair of 2-component rotation eigenvectors, can be associated with a $2 \times 2$ matrix representation of rotations in a Euclidean space without disturbing the plane wave's space-time properties. In a space with more than four dimensions, only rotations in a 4d subspace can be represented. So far all is well known. Now consider that unrepresented rotations do not have eigenvectors, do not make plane waves, and do not contribute phase. The particles e, $\pi,$ and p are assigned rotations in a 4d subspace of 16d, rotations in an 8d subspace of 12d, and rotations in a 12d subspace of 12d, respectively. The electron 4d subspace, assumed to be as likely to align with any one 4d subspace as with any other, produces phase when aligned with the represented 4d subspace in 16d. Similarly, we calculate the likelihood that a 4d subspace of the pion's 8d space aligns with the represented 4d subspace in 12d. The represented 4d subspace is contained in the proton's 12d space, so the proton always acquires phase. By the relationship between mass and phase, the resulting particle phase ratios are the particle mass ratios and these are coincident with the measured mass ratios, within about one percent.

1999 PACS number(s): 03.65.Fd

			\end{abstract}
\pagebreak

\section{Introduction} 

	The frequency with which the wave function of a free particle at rest oscillates is the particle's mass, aside from a conversion factor to accomodate mass and frequency units. Single particle wave functions must transform properly to reflect the symmetries of space-time: Spin 1/2 wave functions are composed of pairs of the 2-component vectors in the $2\times 2$ matix represention of the group of boost and rotation symmetries of space-time. This paper relates the frequencies and masses of the electron, pion, and proton to the properties of $2 \times 2$ representations.

	There is a rule in group theory that a subgroup is more general than the full group. For example, three-sided symmetry in a plane includes all the figures that have six-sided symmetry and more. The rotation group is a subgroup of the Lorentz group, and recent work [1] has shown that the Lorentz-based Dirac equation is a special case of a more general equation true for all pairs of rotation eigenvectors.

	As discussed in Ref.~1 and briefly here in Sec.~2, a spin 1/2 plane wave is completely characterized as a pair of rotation eigenvectors, thereby eliminating the need to consider Lorentz boosts. This means that we can consider the rotation group independently of space-time symmetries and discuss the rotations in a Euclidean space without compromising the space-time properties of the wave functions.

	Rotation eigenvectors naturally gain phase when the rotation angle increases. This gives a nice interpretation of the frequency of a spin 1/2 wave function: the frequency indicates a rate of rotation. In summary, previous work has shown that, given a rotation represented by $2 \times 2$ matrices, pairs of the rotation's eigenvectors make spin 1/2 plane waves and the rotation rate is proportional to the frequency of the plane wave and therefore proportional to particle mass.

	The new idea in this paper is that there may be some rotations that are not represented, do not have eigenvectors, and do not contribute to particle phase. The masses of two otherwise equal mass particles can differ if one particle has more represented rotations than the other.  

	Represented and non-represented rotations appear when one considers $2 \times 2$ matrix representations of rotations in more than four dimensions. In Sec.~4, we discuss the fact that there are two $2 \times 2$ matrix representations in a four dimensional (4d) Euclidean space, while there are no $2 \times 2$ matrix representations for the rotation group in a Euclidean space with five or more dimensions. Therefore in higher dimensional Euclidean spaces, only rotations in 4d subspaces can be represented by $2 \times 2$ matrices.

	In Sec.~5 a sixteen dimensional space is considered. Rotations in two disjoint, special, 4d subspaces, called $2 \times 2$ represented spaces $S_{2 \times 2}$ and $\tilde{S}_{2 \times 2}$, are represented by the two inequivalent $2 \times 2$ representations. Then the electron is associated with rotations in a randomly oriented 4d subspace that on occasion aligns with one of the represented spaces $S_{2 \times 2}$. Only during alignment can the electron's 4d subspace have represented rotations and have eigenvectors whose eigenvalues determine quantum phase. Hence the electron has less mass than is potentially available because its associated rotations are not always represented.   

	Next, rotations in a randomly oriented 8d subspace are associated with a pion. A 4d subspace of the 8d space occasionally aligns with the second represented space $\tilde{S}_{2 \times 2}$ in the 12d subspace not including $S_{2 \times 2}.$ The pion acquires quantum phase whenever any 4d subspace aligns with $\tilde{S}_{2 \times 2}.$ Two factors play a role, the alignment of 4d with 4d in 12d and the number of ways of choosing 4d in 8d. The product of the factors is coincident with the pion-proton mass ratio.

	Having considered 4d and 8d subspaces it remains to consider a randomly oriented 12d subspace. The rotations in the 12d space are associated with a proton. But there are only 12 dimensions in the subspace of the 16d space that doesn't include $S_{2 \times 2},$ hence $\tilde{S}_{2 \times 2}$ is always aligned with some 4d subspace of the arbitrarily oriented 12d subspace. Thus the proton has the maximum available mass because the rotations in some 4d subspace are always represented and are always acquiring phase.

	The model explains the e-$\pi$-p masses as potentially equal particle masses restricted by the need to make spin 1/2 plane waves from eigenvectors of rotations. The plane wave for rotations in 4d give the electron wave function. The number of dimensions for the pion and proton agrees with the quark model: room for two spin 1/2 particles in 8d for the pion and room for three spin 1/2 particles in 12d for the proton. However, the pion and proton in the quark model have two and three viable quarks at all times, whereas here only one of the 4d subspaces in the 8d pion space and in the 12d proton space can have represented rotations at any one time.

\section{Rotation Angle and Quantum Phase} 

	In this section some aspects of a previous paper `Rotation Eigenvectors and Spin 1/2' (RES) are discussed that bear on the problem of quantum phase and mass. 

	In RES [1] a complete set of spin 1/2 wave functions with special rotation properties is obtained, with each 4-spinor of the set is a pair of two-component rotation eigenvectors. Technically these paired 2-component eigenvectors could be called 4-spinors, but the term `4-spinor' brings to mind the Lorentz group and can cause confusion. Here not boosts but exclusively rotations are being represented. 

	As a plausibility argument, consider that in a 3+1 space-time with ``4'' indicating time, the 12 plane and the 34 plane are independent. Hence 12 rotations commute with 34 boosts and, because the number of $2 \times 2$ generators is limited, the eigenvectors of one are eigenvectors of the other. Free particle spin 1/2 wave functions can be written in many equivalent ways and in some, so-called `chiral representations,' the spin is lined up with the motion. The 4-spinor for constant speed in the `3' direction is then a pair of 34 eigenvectors which are simultaneously 12 eigenvectors. The duplication allows us to characterize the 4-spinor for a constant velocity as a pair of rotation eigenvectors. 

	The properties of rotations imply that every pair of rotation eigenvectors satisfies an equation that contains, as a special case, the Dirac equation in momentum representation. Half the rotation angle $\theta$ in the eigenvalue $e^{i\theta/2}$ for a given eigenvector pair is the quantum phase $\phi$ in that pair's constant momentum state of a spin 1/2 particle:
\begin{equation}
\phi = \frac{\theta}{2} = \cosh{(u)} \, t - \sinh{(u)} n^{k} x^{k},
\end{equation}
where the velocity of the particle in space-time $(x^{k},t)$ has magnitude $\tanh{(u)}$ in the  direction of the unit vector $n^{k}$ perpendicular to the rotation plane. The speed $\tanh(u)$ depends on the ratio $e^{u}$ of the paired eigenvectors. It is important to realize that space-time does not rotate when the Euclidean space rotates; rather rotations in Euclidean space provide the phase angle of spin 1/2 plane waves, see Fig.~1.
\begin{figure}[h] 	
\vspace{0in}
\hspace{0.75in}\includegraphics[0,0][360,180]{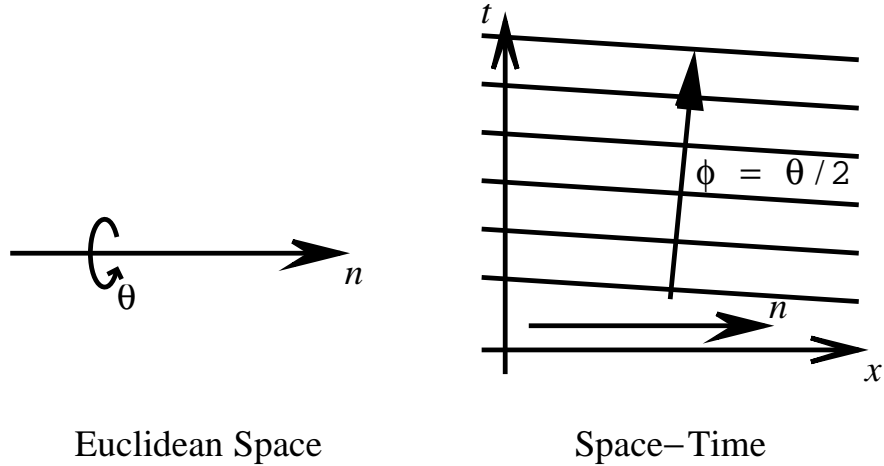}
\caption{Space-time does not rotate when the Euclidean space rotates. As the rotation proceeds about $n,$ the rotation angle $\theta$ increases. In space-time the phase angle $\phi$ = $\theta/2$ increases for the plane wave as the rotation proceeds in the Euclidean space. }
\end{figure}

	Clearly, the right-most member in (1) is a scalar product with signature $(+,-,-,-).$ Therefore, $\phi$ and $\theta$ are invarient under boosts and rotations in space-time $(x^{k},t)$ when $\cosh(u)$ and $\sinh(u) n^{k}$ transform properly. One can choose a boost in the direction of $-n^{k}$ so that the particle's speed vanishes, $\tanh{(u)} \rightarrow 0.$ In this ``rest frame'' the eigenvector ratio collapses, $e^{u} \rightarrow$ 1, and we have $\cosh(u) \rightarrow $ 1, $\sinh(u) \rightarrow $ 0, and
\begin{equation}
\phi = \frac{\theta}{2} =  s ,
\end{equation}
where $s$ is the `proper time,' the time in the rest frame of the particle.

\section{Mass and Phase in Space-time} 

	Radians make sense for the measure of angles $\phi$ and $\theta,$ but proper time $s$ is more commonly measured in seconds or maybe in a distance unit such as meters. It is observed that a heavy particle accumulates a larger phase angle for a given time or distance than a less heavy particle. Hence, mass ``$m$''  enters when the units of $(x^{k},t)$ and $s$ are changed from radians to seconds and meters. Basic quantum mechanics dictates the choice of unit to be
\begin{equation}	
\phi = \frac{\theta}{2} =  m (\frac{s}{m}) = m s^{\prime}.
\end{equation}
We drop the prime on $s^{\prime}$ = $s$ by saying that the unit of $s$ is the `reduced Compton wavelength' $\lambda_{C}/(2 \pi)$ =  $1/m.$ Incorporating Planck's constant $\hbar$ in Joule$\cdot$s and the speed of light $c$ in m/s gives proper time $s$ units in seconds or meters; e.g. for mass $m$ in kilograms and proper time $s$ in seconds one finds that $\phi$ = $m c^{2} s/ \hbar.$

	When the particle is moving with speed $\tanh{(u)}$ as in (1), we can write $(x^{k \, \prime},t^{\prime})$ = $(x^{k}/m,t/m)$. Again we drop the prime by calling $1/m$ the unit for $(x^{k},t).$ The factor $m$ can be combined with the $\cosh{(u)}$ and $\sinh{(u)}$ to define the energy-momentum, 
\begin{equation}	
E \equiv m \cosh{(u)} \hspace{1in} p^{k} \equiv m \sinh{(u)} n^{k}.
\end{equation}
Then it makes sense to say that the mass is the particle's energy in the rest frame. Mass is sometimes characterized by its square which is the Lorentz invarient $E^{2} - \sum_{k}{p^{k}}^{2}$ = $m^2.$ In this paper an alternate characterization must be used.

	We see that mass sets the scale in space-time, so that the quantities $(x^{k},t)$ and $s$ for one particle are the same as those of a second particle. Over a given proper time $s$, the ratio of quantum phases is the same as the mass ratio,
\begin{equation}	
\frac{\Delta \phi_{1}}{\Delta \phi_{2}} = \frac{m_{1} s}{m_{2} s} = \frac{m_{1}}{m_{2}}
\end{equation}
In Sec.~5 we use the ratio of quantum phase as the definition of the mass ratio, because, as discussed in Sec.~2, the phase ratio is the same as the rotation angle ratio $\Delta \theta_{1} / \Delta \theta_{2}$ and we seek to define physical quantities in terms of rotations.

\pagebreak

\section{$2 \times 2$ Representations in Four or More Dimensions} 

	 The electron-pion-proton mass scheme presented in Sec.~5 hinges on the properties of $2 \times 2$ matrix representations of rotations in more than three dimensions. The needed properties are discussed in this section.

	In three dimensions any rotation can be obtained by combining rotations in three coordinate planes, call them planes 12, 23, and 31, where 1,2,3 refer to rectangular coordinates $x^{i},$ $i\in \{1,2,3\},$ in a Euclidean space $E_{3{\mathrm{{\mathrm{d}}}}}.$ For example, the $2 \times 2$ matrix $R$ = $1 + \sigma_{12} \delta \theta,$ where `1' is the unit $2 \times 2$ matrix and $\sigma_{12}$ is the generator, represents a rotation through a small angle $\delta \theta$ in the 12 plane. Let $\sigma_{12},$ $\sigma_{23},$ and $\sigma_{31}$ be the generators of the $2 \times 2$ rotation matrices representing the rotations in these coordinate planes.  One can show that these three generators do not commute, $\sigma_{12} \sigma_{23} \neq$ $\sigma_{23} \sigma_{12},$ etc. Also any $2 \times 2$ matrix can be written as a linear combination of the three generators together with the unit $2 \times 2$ matrix. For background see references [2] and [3], for example.

	In four dimensional Euclidean space, $E_{4{\mathrm{d}}},$ rotations can also be represented by $2 \times 2$ matrices. To indicate the way this works, consider the 12 and 34 planes. Since these planes are independent, the rotations in these planes commute. Since different spin matrices do not commute, the same spin matrix generates rotations in the 12 and in the 34 planes.

	In more detail, we see that there are only four independent $2 \times 2$ matrices and the unit $ 2 \times 2$ matrix commutes with everything, so the generator $\sigma_{34}$ can be restricted to a linear combination of $\sigma_{12},$ $\sigma_{23},$ and $\sigma_{31},$ i.e. $\sigma_{34}$ = $\alpha \sigma_{12}$ + $\beta \sigma_{23}$ + $\gamma \sigma_{31},$ with real valued coefficients. Now the independence of rotations in planes 12 and 34 implies that $\sigma_{12} \sigma_{34}$ = $\sigma_{34} \sigma_{12}$ and substituting the expanded form of $\sigma_{34}$ gives $\beta$ = $\gamma$ = 0 while $\alpha$ is undetermined except that $\alpha$ cannot be zero. Normalizing the generator, $\sigma_{34}^{\dagger} \sigma_{34}$ = 1 forces $\alpha^2$ to be unity, $\alpha^2$ = 1. We can choose $\alpha$ to be either +1 or $-1$ and we find that $\sigma_{34}$ = $\pm \sigma_{12},$ so there are two $ 2 \times 2$ matrix representations of the rotations in four dimensions.

	While we can fit $2 \times 2$ representations to the rotation group in four dimensions, five dimensions are too many. Consider the 35 plane. Rotations in coordinate plane 35 are independent of rotations in 12, but 35 and 34 rotations are not independent. Therefore, the 35 generator would need to commute with the 12 generator, but not commute with the 34 generator. As just shown the 12 and 34 generators are the same within a sign, hence there can be no 35 generator and rotations in 5 or more dimensions do not have a $2 \times 2$ matrix representation.

	The conclusions we need below are that there are two $2 \times 2$ representations of the rotation group in four dimensions and that there are no $2 \times 2$ representations of rotations in five or more dimensions.

\section{The electron-pion-proton mass ratio} 

	In this section a way of explaining the electron-pion-proton mass ratio is presented. The explanation combines facts with rules to give conclusions consistent with known results. There is, of course, a great deal of flexibility here: one can choose to highlight some facts and ignore others and one can make up the rules to fit desired conclusions. On the other hand I know of only two attempts at explaining the electron-proton mass ratio, one a completely different approach by Eddington [4] and a second [5] that is along the lines of, but cruder than, the explanation here.
	
	The `facts' needed for the explanation of the e-$\pi$-p mass ratio are collected in three propositions:

	I. The pair of rotation eigenvectors that make up a constant velocity free particle spin 1/2 plane wave accumulates quantum phase $\phi$ as the rotation angle $\theta$ increases, $\phi$ = $\theta/2,$ as discussed in Sec.~2 and illustrated in Fig~1. 

	II. There are two inequivalent $2 \times 2$ representations of the rotation group in four dimensions and there are no $2 \times 2$ representations of rotations in five or more dimensions. See Sec.~4.

	III. The electron, pion, and proton masses $m_{e},$ $m_{\pi},$ and $m_{p}$ satisfy two numerical coincidences, accurate to a couple of percent,
\begin{equation} 
\frac{m_{e}}{m_{p}} \approx \frac{4! 12!}{16!} \hspace{1in} \frac{m_{\pi}}{m_{p}} \approx 
\frac{4! 8!}{12!} \times \frac{8!}{4! 4!},
\end{equation}
where we can use the values $m_{e}$ = 0.511 MeV, $m_{\pi}$ = 135 meV, and $m_{p}$ = 938 MeV which are rounded from the more accurate values given in a recent Particle Data Group compilation [6]. We turn now to the explanation.

	Consider a sixteen dimensional Euclidean space $E_{16{\mathrm{d}}}$. By II, the full rotation group in sixteen dimensions does not have a $2 \times 2$ representation. But rotations confined to a four dimensional subspace can be given a $2 \times 2$ matrix representation. There are two inequivalent $2 \times 2$ representations and we select two 4d subspaces, calling them $2 \times 2$ represented spaces $S_{2 \times 2}$ and $\tilde{S}_{2 \times 2}$ for \underline{\textit{S}}pinor represented spaces, see Fig.~2a. Rotations in $S_{2 \times 2}$ are represented by one $2 \times 2$ representation and rotations in $\tilde{S}_{2 \times 2}$ are represented by the other. Up to four such represented spaces could be allowed; the model works with these two. The $2 \times 2$ spaces $S_{2 \times 2}$ and $\tilde{S}_{2 \times 2}$ are required to obey the following rule:

	Rule 1. The $2 \times 2$ represented spaces $S_{2 \times 2}$ and $\tilde{S}_{2 \times 2}$ are disjoint. Thus we allow $S_{2 \times 2}$ to be any 4d subspace of $E_{16{\mathrm{d}}}$ while we require that $\tilde{S}_{2 \times 2}$ must be a 4d subspace in the remaining 12d subspace $H_{12{\mathrm{d}}}$ = $E_{12{\mathrm{d}}}-S_{2 \times 2}.$
\begin{figure}
\vspace{0in}
\hspace{0.4in}\includegraphics[0,0][414,180]{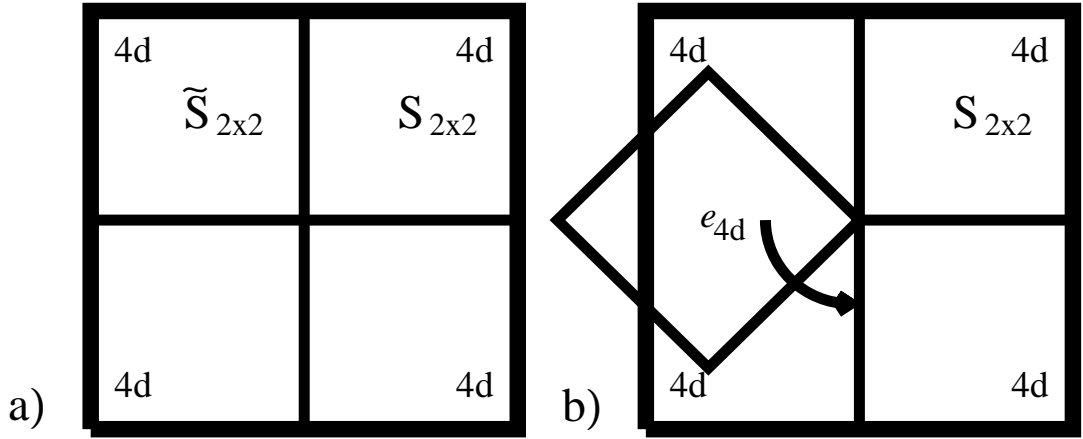}
\caption{a) 16-dimensional Euclidean space $E_{16{\mathrm{d}}}$ with two {\underline{\textit{S}}}pinor represented spaces $S_{2 \times 2}$ and $\tilde{S}_{2 \times 2},$ in which rotations are represented by $2 \times 2$ matrices. b) Rotations in a 4d subspace $e_{4{\mathrm{d}}}$ are represented only when $e_{4{\mathrm{d}}}$ aligns with $S_{2 \times 2}.$ }
\end{figure}

	By I, a rotation by an angle $\theta$ in any plane of $S_{2 \times 2}$ changes the phase of that rotation's eigenvectors by $\theta.$ Since $\phi$ = $\theta/2$ the quantum phase of the particle described by the paired eigenvectors of $S_{2 \times 2}$ increases by $\theta/2$. Similar remarks hold for $\tilde{S}_{2 \times 2}.$ 

	In contrast, a rotation that is not represented has neither eigenvectors nor eigenvalues. An unrepresented rotation generates no phase. These comments bring us to the second rule:

	Rule 2. The rotations in a 4d subspace $w_{4{\mathrm{d}}}$ of $E_{16{\mathrm{d}}}$ can be associated with a particle. But a given rotation in $w_{4{\mathrm{d}}}$ contributes to the particle's phase a fraction of the phase for a similar rotation in $S_{2 \times 2}.$ The fraction is determined by the likelihood, defined below, that $w_{4{\mathrm{d}}}$ is oriented parallel to the $2 \times 2$ represented space $S_{2 \times 2}.$

	What is the likelihood that two 4d subspaces are aligned? We proceed by analogy with the polarization of light. There are an infinite number of possible polarizations, just as there are an infinite number of 4d subspaces. But any polarization state is a linear composition of just two independent states, and we can take these two states to be linear polarized in orthogonal directions. 

	Call two 4d subspaces `orthogonal' if any vector in one is perpendicular to all the vectors in the other. The idea of orthogonal 4d subspaces is developed in the Appendix where an orthonormal basis of 4d subspaces is defined and discussed. Let $s_{\alpha},$ where $\alpha \in$ $\{1, 2, ..., 1820\},$ be an orthonormal basis of 4d subspaces in $E_{16{\mathrm{d}}}$. Note that, by (17) with $N$ = 16, the number of different subspaces in $E_{16{\mathrm{d}}},$ $16!/(4!12!)$ = 1820, is the same as the number of ways of picking four unit vectors from the sixteen vectors in an orthonormal basis of $E_{16{\mathrm{d}}}.$

\pagebreak

	\underline{Electron.} Let $e_{4{\mathrm{d}}}$ be a 4-d subspace of $E_{16{\mathrm{d}}}.$ We choose `$e_{4{\mathrm{d}}}$' because this 4d subspace turns out to be associated with an {\underline{\textit{e}}}lectron. Suppose the orientation of $e_{4{\mathrm{d}}}$ in $E_{16{\mathrm{d}}}$ is either random or circular so that $e_{4{\mathrm{d}}}$ is as likely to be in any one orientation as in any other, see Fig.~2b.  
	 
	By (18), we may write the 4d subspace $e_{4{\mathrm{d}}}$ as a linear combination of the basis spaces:
\begin{equation} 
e_{4{\mathrm{d}}} = \sum_{\alpha} e_{\alpha} s_{\alpha},
\end{equation}
where, as shown in the Appendix, $e_{\alpha}$ is the determinent of the $4 \times 4$ matrix formed by the sixteen scalar products of the four vectors in an orthonormal basis in $e_{4{\mathrm{d}}}$ and the four vectors in an orthonormal basis in $s_{\alpha}.$ By (19), the squares of the coefficients add to unity,
\begin{equation} 
\sum_{\alpha} e_{\alpha}^2 = 1,
\end{equation}
see the Appendix for details. Since the sum is unity and by analogy with discussions of linear polarization, we can introduce probability language. Define $e_{\alpha}^{2}$ to be the `likelihood that the 4d subspace $e_{4{\mathrm{d}}}$ aligns with the basis subspace $s_{\alpha}.$'

Since the orientation of $e_{4{\mathrm{d}}}$ is random or circular, the coefficients $e_{\alpha}$ are all equal, on average. Thus the likelihood that the 4d subspace $e_{4{\mathrm{d}}}$ aligns with one of the basis subspaces $s_{1}$ is the inverse of the number of subspaces in the basis, hence
\begin{equation} 
 ({\mathrm{The}} \:{\mathrm{likelihood}} \:{\mathrm{of}} \: 4{\mathrm{d}} \: {\mathrm{aligning}} \: {\mathrm{with}} \: 4{\mathrm{d}} \: {\mathrm{in}} \: {\mathrm{16{\mathrm{d}}}}) = \langle e_{\alpha}^{2} \rangle = \frac{4! 12!}{16!} = \frac{1}{1820}.
\end{equation}

	By Rule 2, rotations in $e_{4{\mathrm{d}}}$ contribute phase factors to eigenvectors only when the rotations are represented and that happens only when $e_{4{\mathrm{d}}}$ aligns with $S_{2 \times 2}.$ For simplicity choose $S_{2 \times 2}$ to be one of the basis 4d subspaces, say $S_{2 \times 2}$ = $s_{1}.$ Then the phase increase $\Delta \phi_{e}$ for rotations in $e_{4{\mathrm{d}}}$ is a fraction of the phase increase $\Delta \phi_{S}$ for rotation eigenvectors in $S_{2 \times 2},$ $\Delta \phi_{e}$ = $(4! 12!/16!)\Delta \phi_{S}.$ By (5), the ratio of phase change is equivalent to a mass ratio, hence
 \begin{equation} 
 \frac{M_{e}}{M} = \frac{\Delta \phi_{e}}{\Delta \phi_{S}} = \frac{4! 12!}{16!} = \frac{1}{1820},
\end{equation}
where $M_{e}$ and $M$ are the masses of the particle described by paired eigenvectors in $e_{4{\mathrm{d}}}$ and $S_{2 \times 2}$, respectively. 

 	During the calculation of (10) we have ignored the possibility that the other $2 \times 2$ represented space $\tilde{S}_{2 \times 2}$ aligns with $e_{4{\mathrm{d}}}.$ In fact we impose yet another rule:

Rule 3. The $2 \times 2$ represented space $S_{2 \times 2}$ is reserved for 4d subspaces, i.e. electrons. The other $2 \times 2$ represented space $\tilde{S}_{2 \times 2}$ takes care of the pion and proton. 

	\underline{Pion and Proton.} We do not consider aligning a 4-d subspace like $e_{4{\mathrm{d}}}$ with $\tilde{S}_{2 \times 2}$ because such considerations seem to lead to no new useful numerical coincidences. One might expect to repeat the result (10) except that Rule 1 implies that $\tilde{S}_{2 \times 2}$ can align only with 4d subspaces in the $16-4$ = 12d space $H_{12{\mathrm{d}}}$. Furthermore we are not interested in a 4d subspace because, in the quark model, a pion consists of two spin 1/2 quarks and a proton has three spin 1/2 quarks.
\begin{figure}
\vspace{0in}
\hspace{0.4in}\includegraphics[0,0][415,180]{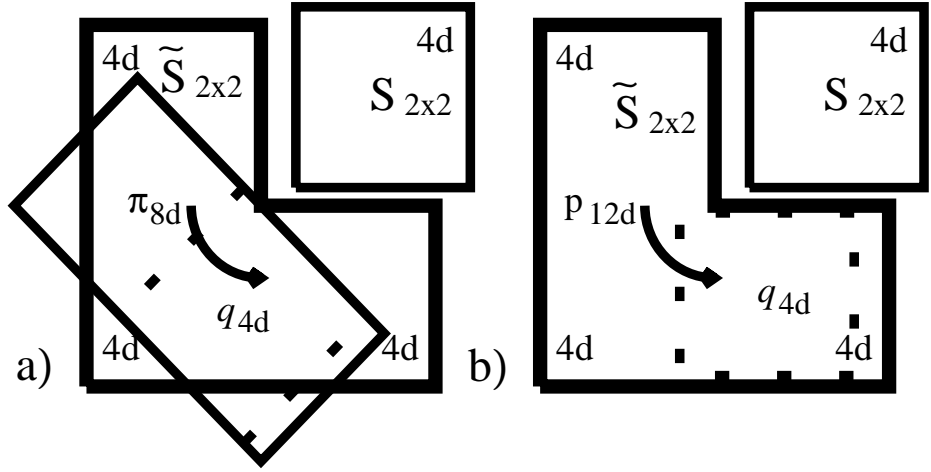}
\caption{a) Rotations in a 4d subspace $q_{4{\mathrm{d}}}$ of an 8d space $\pi_{8{\mathrm{d}}}$ in $H_{12{\mathrm{d}}}$ = $E_{16{\mathrm{d}}}-S_{2 \times 2}$ are represented upon alignment with $\tilde{S}_{2 \times 2}.$ b) The 12d space $p_{12{\mathrm{d}}}$ differs from $H_{12{\mathrm{d}}}$ only in its orientation. Thus some 4d subspace of $p_{12{\mathrm{d}}}$ is always represented, though not the $q_{4{\mathrm{d}}}$ shown.}
\end{figure}

	Consider a 4 + 4 = 8d subspace $\pi_{8{\mathrm{d}}}$ of the 12d space $H_{12{\mathrm{d}}}$ = $E_{16{\mathrm{d}}}-S_{2 \times 2},$ see Fig.~3a. Obviously not all  eight dimensions of $\pi_{8{\mathrm{d}}}$ can align with $\tilde{S}_{2 \times 2}$ at once. So we choose a 4-d subspace $q_{4{\mathrm{d}}}$ of $\pi_{8{\mathrm{d}}}$ and ask for the likelihood that $q_{4{\mathrm{d}}}$ is oriented parallel with $\tilde{S}_{2 \times 2}$ in the 12-d space $H_{12{\mathrm{d}}}.$ Expand the 4d subspace $q_{4{\mathrm{d}}}$ as a linear combination of basis 4d subspaces $s_{\beta}^{\prime}$ in 12-dimensions, $q_{4{\mathrm{d}}}$ = $\sum_{\beta}q_{\beta} s_{\beta}^{\prime},$ as in (7). There are fewer basis 4d subspaces in 12d than in 16d; the number in 12d is $12!/(4!8!)$, by (17) with $N$ = 12. Since $q_{4{\mathrm{d}}}$ is randomly or circularly oriented in $H_{12{\mathrm{d}}},$ all basis 4d subspaces $s_{\beta}^{\prime},$ $\beta \in$ $\{1,...,12!/(4!8!),$ are equally likely and we have
\begin{equation} 
 ({\mathrm{The}} \: {\mathrm{likelihood}} \: {\mathrm{of}} \: 4{\mathrm{d}} \: {\mathrm{aligning}} \: {\mathrm{with}}  \: 4{\mathrm{d}} \: {\mathrm{in}} \: {\mathrm{12{\mathrm{d}}}}) = \langle q_{\beta}^{2} \rangle = \frac{4! 8!}{12!} = \frac{1}{495}.
\end{equation}

	The probability of something happening is the probability for one way that the thing can happen multiplied by the number of similar ways. Hence we need to know the number of ways of choosing a 4-d subspace $q_{4{\mathrm{d}}}$ of $\pi_{8{\mathrm{d}}}.$ We take the number of independent choices to be the number of basis 4d subspaces in $\pi_{8{\mathrm{d}}},$ i.e. $8!/(4!4!)$ by (17) with $N$ = 8,
 \begin{equation} 
 ({\mathrm{The}} \:{\mathrm{number}} \: {\mathrm{of}}  \: {\mathrm{distinct}}  \: 4{\mathrm{d}} \: {\mathrm{subspaces}} \: {\mathrm{in}} \:{\mathrm{8{\mathrm{d}}}}) = \frac{8!}{4!4!} = 70.
\end{equation}
By Rule 2, the rate at which some rotation in $\pi_{8{\mathrm{d}}}$ gains phase is the likelihood that some 4d subspace $q_{4{\mathrm{d}}}$ aligns with $\tilde{S}_{2 \times 2}$ multiplied by the number of distinct 4d subspaces. Hence the increase in phase for the 8d space $\pi_{8{\mathrm{d}}}$ is a fraction of the increase of phase for $\tilde{S}_{2 \times 2}.$ By (5) that fraction would be seen as the mass ratio,
 \begin{equation} 
 \frac{M_{\pi}}{M} = \frac{\Delta \phi_{\pi}}{\Delta \phi_{S}} = \frac{4! 8!}{12!} \times \frac{8!}{4!4!} = \frac{70}{495},
\end{equation}
where $M_{\pi}$ is the mass of the particle described by the rotations in $\pi_{8{\mathrm{d}}}.$  

	The final subspace considered here is a $4 + 4 + 4$ = 12d subspace $p_{12{\mathrm{d}}}$ of the 12d space $H_{12{\mathrm{d}}},$ named for {\underline{\textit{H}}}adron space, see Fig.~3b. Clearly $p_{12{\mathrm{d}}}$ and $H_{12{\mathrm{d}}}$ can differ only in their relative orientation. The likelihood that a randomly oriented 4d subspace $q_{4{\mathrm{d}}}$ of $p_{12{\mathrm{d}}}$ aligns with $\tilde{S}_{2 \times 2}$ has already been found, see (11). We also need the number of ways of choosing a 4d subspace $q_{4{\mathrm{d}}}$ in the 12d space $p_{12{\mathrm{d}}}$ and that is the number of basis 4d subspaces. Thus, by (17) with $N$ = 12,
 \begin{equation} 
 ({\mathrm{The}} \: {\mathrm{number}} \: {\mathrm{of}}  \: {\mathrm{distinct}}  \: 4{\mathrm{d}} \: {\mathrm{subspaces}} \: {\mathrm{in}} \:{\mathrm{12{\mathrm{d}}}})  = \frac{12!}{4!8!} = 495.
\end{equation}
The rate at which some rotation eigenvector for a rotation in $p_{12{\mathrm{d}}}$ gains phase is the likelihood that some 4d subspace $q_{4{\mathrm{d}}}$ aligns with $\tilde{S}_{2 \times 2}$ multiplied by the number of such spaces. Hence the mass ratio is 
 \begin{equation} 
 \frac{M_{p}}{M} = \frac{\Delta \phi_{p}}{\Delta \phi_{S}} = \frac{4! 8!}{12!} \times \frac{12!}{4!8!} = 1,
\end{equation}
where $M_{p}$ is the mass of the particle described by the rotations in $p_{12{\mathrm{d}}}.$ This result is obvious since $\tilde{S}_{2 \times 2}$ is contained in $H_{12{\mathrm{d}}}$ which is the same space as $p_{12{\mathrm{d}}}$ aside from orientation.

	\underline{Discussion.} Comparing (8), (11), and (13) with the numerical coincidences (6) we identify $M_{e},$ $M_{\pi},$ and $M_{p}$ with the masses $m_{e},$ $m_{\pi},$ and $m_{p}$ of the electron, pion, and proton, respectively. The values of the mass $M$ inferred for the represented spaces $S_{2 \times 2}$ and $\tilde{S}_{2 \times 2}$ from (8), (11), and (13) are 930 MeV, 955 MeV, and 938 MeV, respectively. As a gauge of the model's accuracy note that these values are within a couple of percent of their average value, 941 MeV. 

	One should notice that the coefficients in a general linear combination $\sum_{\alpha} v_{\alpha} s_{\alpha}$ of basis 4d subspaces $s_{\alpha}$ need not have the special form required of coefficients $e_{\alpha}$ and $w_{\alpha}$ in (7) and (18) for the 4d subspaces $e_{4d}$ and $w_{4d}.$ Indeed we could choose an unrestricted sum in place of (7) for an electron, as long as the square of the coefficients all have the same average, 1/1820. An electron state with rotations in a 4d subspace is only required when that state aligns with $S_{2 \times 2}.$ 

	The treatment of quarks in this model differs from that in the conventional quark model. In the conventional model, both quarks in a pion have mass at all times. In the just-described model when a 4d subspace $q_{4{\mathrm{d}}}$ of $\pi_{8{\mathrm{d}}}$ does align with the $\tilde{S}_{2 \times 2}$ filter, just one such space aligns, not two. Both 4d subspaces are required to make the pion's 8d space $\pi_{8{\mathrm{d}}}$, but only the rotations in one 4d subspace can contribute to the phase, and thereby supply mass, at any given time. 

	Likewise the 12d space for the proton has rotations in enough dimensions to describe three independent spin 1/2 particles and that is consistent with the quark model. The calculation of the mass ratio (13) hinges on having some 4d subspace always aligned with $\tilde{S}_{2 \times 2}$ and contributing phase by rule 2. Rotations in the 8d space $p_{12{\mathrm{d}}} - q_{4{\mathrm{d}}}$ do not contribute phase. One might say simply that one quark has mass while the other two do not have mass, at any given instant of time. However there is a complication. Any 4d subspace in 12d is the superposition of a population of $12!/(4!8!)$ = 495 distinct 4d subspaces, each of which is as likely to be the aligned subspace as any of the others. So the represented rotations in some 4d space give a proton a continuous supply of mass at all times, but those represented 4d spaces do not make one quark, nor three quarks taking turns, but rather 495 distinct subspaces each with some likelihood of being represented.

\appendix
\section{Orthonormal 4d subspaces in Euclidean Spaces}

	The vectors discussed in this Appendix should not be confused with the rotation eigenvectors discussed in the main part of the paper. The eigenvectors there have 2-components since they occur with a $2 \times 2$ matrix representation of the rotation group in at most four dimensions. The vectors in this Appendix have $N$-components and are associated with the ordinary $N \times N$ matrix representation of rotations in $N$-dimensional Euclidean space.

	Let $E_{N{\mathrm{d}}}$ be an $N$-dimensional Euclidean space with $N \geq$ 4. Let $\{E_{1}, ..., E_{N}\}$ be an orthonormal basis for $E_{N{\mathrm{d}}},$ i.e. $E_{I} \cdot E_{J}$ = $\delta_{IJ},$ where $\delta_{IJ}$ is 1 when $I = $ $J,$ $\delta_{IJ}$ is 0 otherwise, and $I, J \in$ $\{1, ..., N\}.$ The scalar product of two vectors, $v_{1}$ and $v_{2}$ can be written as $v_{1} \cdot v_{2}$ = $ \sum_{J} (v_{1} \cdot E_{J}) (v_{2} \cdot E_{J})$ = $\sum_{J} v_{1}^{J} v_{2}^{J}.$  

	Select any two 4d subspaces $w_{4{\mathrm{d}}}$ and $w_{4{\mathrm{d}}}^{\prime}$ each with an orthonormal basis $ w_{i}$ and $ w_{i}^{\prime},$ where $i \in$ $\{1,2,3,4\}.$ The basis of $w_{4{\mathrm{d}}}$ can be written as $w_{i}$ = $a_{iJ} E_{N},$ in terms of the $4 \times N$ matrix $a_{iJ} \equiv$ $w_{i} \cdot E_{J}.$ There is a matrix $a_{iJ}^{\prime}$ for $w_{4{\mathrm{d}}}^{\prime}$ defined in the same way.

	Define the scalar product of the two 4d subspaces, $w_{4{\mathrm{d}}}$ and $w_{4{\mathrm{d}}}^{\prime}$, to be the determinent of the product of the matrix $a_{iJ}$ with the transpose of $a_{iJ}^{\prime}$, $w_{4{\mathrm{d}}} \cdot w_{4{\mathrm{d}}}^{\prime} \equiv$ $\det{(\sum_{J}a_{iJ} a_{kJ}^{\prime}})$ = $\det{(w_{i} \cdot w_{k}^{\prime})}.$ We use a center dot to indicate both the scalar product of two vectors and the scalar product of two 4d subspaces. 

	If a vector $V$ in $w_{4{\mathrm{d}}}$ is perpendicular to all vectors in $w_{4{\mathrm{d}}}^{\prime},$ then we can realign the basis $w_{i}$ so that $w_{1}$ is in the direction of $V.$ In that case the matrix $w_{i} \cdot w_{k}^{\prime}$ has a column of zeros, $w_{1} \cdot w_{k}^{\prime}$ = $V \cdot w_{k}^{\prime}$ = 0, and the determinant is zero. Thus the two 4d subspaces, $w_{4{\mathrm{d}}}$ and $w_{4{\mathrm{d}}}^{\prime}$, are `orthogonal' if any vector in one is orthogonal to all vectors in the second. 

	Next we define an `orthonormal basis' for the 4d subspaces of $E_{N{\mathrm{d}}}$ to be a set of 4d subspaces $s_{\alpha},$ such that 
\begin{equation}
s_{\alpha} \cdot s_{\beta} = \delta_{\alpha \beta}.
\end{equation}
One choice of an orthonormal basis for the 4d subspaces of  $E_{N{\mathrm{d}}}$ is a collection of 4d subspaces spanned by four suitably chosen basis vectors of $E_{N{\mathrm{d}}}.$ Let $\{E_{A}, E_{B}, E_{C}, E_{{\mathrm{d}}}\}$ denote a typical basis for one of the 4d subspaces $s_{\alpha}.$ Clearly each set of integers $\{A,B,C,D\}$ must be chosen from among $\{1,2,...,N\}$ and no two integers in a set can be equal. Orthogonality means that no two sets can contain the same four integers.  Hence the order of integers in the set $\{A,B,C,D\}$ is not important; two sets containing the same four integers but in different order determine the same 4d subspace. Thus the number of 4d subspaces in a basis is the same as the number of ways four integers can be chosen from $\{1, ..., N\}$ irrespective of the order in which they are chosen, i.e.  
\begin{equation} 
 ({\mathrm{The}} \:{\mathrm{number}} \:{\mathrm{of }} \:{\mathrm{4{\mathrm{d}}}} \:{\mathrm{subspaces}} \: {\mathrm{in}} \:{\mathrm{the}} \:{\mathrm{basis}} \: s_{\alpha}) = \frac{N(N-1)(N-2)(N-3)}{4!} = \frac{N!}{4! (N-4)!}. 
\end{equation}

	Any 4d subspace, say $w_{4{\mathrm{d}}},$ can be written as a linear combination of the basis 4d subspaces $s_{\alpha},$ 
\begin{equation}
w_{4{\mathrm{d}}} = \sum_{\alpha} w_{\alpha} s_{\alpha},
\end{equation}
 where $w_{\alpha}$ = $w_{4{\mathrm{d}}} \cdot s_{\alpha}.$ The equality in (18) means that the scalar product of any 4d subspace $w_{4{\mathrm{d}}}^{\prime}$ with the left side, $w_{4{\mathrm{d}}}\cdot w_{4{\mathrm{d}}}^{\prime},$ is the same as the scalar product of $w_{4{\mathrm{d}}}^{\prime}$ with the right side, $(\sum_{\alpha} w_{\alpha} s_{\alpha}) \cdot w_{4{\mathrm{d}}}^{\prime}.$ Since $ \sum_{\alpha} w_{\alpha}^2$ = $w_{4{\mathrm{d}}} \cdot w_{4{\mathrm{d}}}$ = $\det (w_{i} \cdot w_{k}) $ = $\det (\delta_{ik})$ = 1, the sum of the squares of the coefficients in (18) is unity, 
\begin{equation}
 \sum_{\alpha} w_{\alpha}^2 = 1.
\end{equation}

	A linear combination of basis 4d subspaces, $v \equiv$ $\sum_{\alpha} v_{\alpha} s_{\alpha},$ with arbitrary scalar coefficients $v_{\alpha}$ has a well defined scalar product with any other linear combination. For example the scalar product of $v$ and $w_{4d}$ is just $v \cdot w_{4d}$ = $\sum_{\alpha} v_{\alpha} w_{\alpha},$ since the $s_{\alpha}$ are orthonormal, as in (16). The set of all such $v$s makes a vector space, though not all these $v$s represent 4d subspaces.

\pagebreak

\section{Problems}

1. Show that the frequencies $\nu$ listed in the abstract follow from the formula $\nu$ = $m c^2/ h.$

\noindent 2. $2 \times 2$ matrices can represent rotations in 3-dimensional Euclidean spaces. Define orthogonal 3d subspaces in the same way orthogonal 4d subspaces are defined in the Appendix. Show that the number of orthonormal 3d subspaces in $3a$-dimensional space is the binomial coefficient $(3a)!/[3! (3a-3)!].$ Compare the values for $a \in$ $\{1,2,3,4\}$ with mass ratios of various stable, or at least long-lived, particles. 

\noindent 3. By realigning the orthonormal basis in the 4d subspace $w_{4{\mathrm{d}}}$ discussed in the Appendix, we obtain a second orthonormal basis of $w_{4{\mathrm{d}}}$. What effect does the change of basis have on the subspace scalar product $w_{4{\mathrm{d}}} \cdot w_{4{\mathrm{d}}}^{\prime}$ defined in the Appendix? What happens to $w_{4{\mathrm{d}}} \cdot s_{\alpha}$ upon changing to a new orthonormal basis in the $N$-dimensional space containing the 4d subspaces $w_{4{\mathrm{d}}}$ and $s_{\alpha}$?

\noindent 4. In a chiral representation of free spin 1/2 particle wave functions, show that a boost in the $x$-direction applied to the 4-component wave function for a particle at rest with spin in the $x$-direction multiplies the right-handed 2-spinor by a numerical factor and divides the left-handed 2-spinor by the same factor. Hence the right-and left-handed 2-spinors remain eigenspinors of $x$-rotations. 

\noindent 5. How many of the $nN$ components of an orthonormal basis for an $n$d subspace in an $N$d Euclidean space are arbitrary? For what values of $N$ and $n$ do the coefficients in the sum $v$ = $\sum_{\alpha} v_{\alpha} s_{\alpha},$ with $\sum_{\alpha} v_{\alpha}^{2}$ = 1, suffice to determine an $n$d subspace $v?$ Also discuss the cases (i) $n$ = 2 with $N$ = 3 and (ii) $n$ = 2 with $N$ = 4.


\end{document}